\documentclass[a4paper,fleqn]{cas-dc}\usepackage[]{graphicx}\usepackage[]{color}
\makeatletter
\def\maxwidth{ %
  \ifdim\Gin@nat@width>\linewidth
    \linewidth
  \else
    \Gin@nat@width
  \fi
}
\makeatother

\definecolor{fgcolor}{rgb}{0.345, 0.345, 0.345}
\newcommand{\hlnum}[1]{\textcolor[rgb]{0.686,0.059,0.569}{#1}}%
\newcommand{\hlstr}[1]{\textcolor[rgb]{0.192,0.494,0.8}{#1}}%
\newcommand{\hlopt}[1]{\textcolor[rgb]{0,0,0}{#1}}%
\newcommand{\hlstd}[1]{\textcolor[rgb]{0.345,0.345,0.345}{#1}}%
\newcommand{\hlkwb}[1]{\textcolor[rgb]{0.69,0.353,0.396}{#1}}%
\newcommand{\hlkwc}[1]{\textcolor[rgb]{0.333,0.667,0.333}{#1}}%
\newcommand{\hlkwd}[1]{\textcolor[rgb]{0.737,0.353,0.396}{\textbf{#1}}}%

\usepackage{framed}
\makeatletter
\newenvironment{kframe}{%
 \def\at@end@of@kframe{}%
 \ifinner\ifhmode%
  \def\at@end@of@kframe{\end{minipage}}%
  \begin{minipage}{\columnwidth}%
 \fi\fi%
 \def\FrameCommand##1{\hskip\@totalleftmargin \hskip-\fboxsep
 \colorbox{shadecolor}{##1}\hskip-\fboxsep
     \hskip-\linewidth \hskip-\@totalleftmargin \hskip\columnwidth}%
 \MakeFramed {\advance\hsize-\width
   \@totalleftmargin\z@ \linewidth\hsize
   \@setminipage}}%
 {\par\unskip\endMakeFramed%
 \at@end@of@kframe}
\makeatother

\definecolor{shadecolor}{rgb}{.97, .97, .97}
\definecolor{messagecolor}{rgb}{0, 0, 0}
\definecolor{warningcolor}{rgb}{1, 0, 1}
\definecolor{errorcolor}{rgb}{1, 0, 0}
\newenvironment{knitrout}{}{} 

\usepackage{alltt}
\usepackage[numbers]{natbib}
\usepackage{lineno, blindtext, color}
\usepackage{multicol}
\usepackage{amssymb}
\usepackage{amsmath}
\usepackage{bm}
\usepackage{color}
\usepackage{graphicx}
\usepackage{multirow} 
\usepackage{multirow} 
\usepackage{threeparttable}
\usepackage{lscape}
\def\tsc#1{\csdef{#1}{\textsc{\lowercase{#1}}\xspace}}
\tsc{WGM}
\tsc{QE}
\tsc{EP}
\tsc{PMS}
\tsc{BEC}
\tsc{DE}

\let\code=\texttt
\let\proglang=\textsf
\newcommand{\pkg}[1]{{\fontseries{b}\selectfont #1}}

\IfFileExists{upquote.sty}{\usepackage{upquote}}{}
\begin{document}
\let\WriteBookmarks\relax
\def\floatpagepagefraction{1}
\def\textpagefraction{.001}
\shorttitle{\texttt{ForestFit}}
\shortauthors{Mahdi Teimouri, Jeffrey W. Doser, and Andrew O. Finley}
\title [mode = title]{\texttt{ForestFit}: An R package for modeling tree diameter distributions}
\author[1]{Mahdi Teimouri}[type=editor, auid=000,bioid=1, prefix=,   role=,  orcid=https://orcid.org/0000-0002-5371-9364]
\ead{teimouri@aut.ac.ir}
\address[1]{Department of Statistics, Faculty of Science and Engineering, Gonbad Kavous University, Gonbad Kavous, Iran}
\author[2]{Jeffrey W. Doser}
\author[2]{Andrew O. Finley}
\address[2]{Department of Forestry, Michigan State University, East Lansing, MI, USA}
\cortext[cor1]{Mahdi Teimouri}
\begin{abstract}
Modeling the diameter distribution of trees in forest stands is a common forestry task that supports key biologically and economically relevant management decisions. The choice of model used to represent the diameter distribution and how to estimate its parameters has received much attention in the forestry literature; however, accessible software that facilitates comprehensive comparison of the myriad modeling approaches is not available. To this end, we developed an \proglang{R} package called \pkg{ForestFit} that simplifies estimation of common probability distributions used to model tree diameter distributions, including the two- and three-parameter Weibull distributions, Johnson's SB distribution, Birnbaum-Saunders distribution, and finite mixture distributions. Frequentist and Bayesian techniques are provided for individual tree diameter data, as well as grouped data. Additional functionality facilitates fitting growth curves to height-diameter data. The package also provides a set of functions for computing probability distributions and simulating random realizations from common finite mixture models.
\end{abstract}



\maketitle

\section{Introduction}
The diameter distribution of trees in a stand is a relatively simple measure used to inform forest management decisions that have both biological and economical relevance  \citep{Bailey1973quantifying}. Determining models and associated parameter estimation techniques to characterize an entire stand's diameter distribution based upon a sample of trees has been an important topic in forestry over the last 50 years \citep{green1994bayesian,merganivc2006characterisation}. Using an appropriate model to describe the diameter distribution enables an evaluation of the stand growth as well as potential estimates of stand volume. Numerous probability distributions have been proposed and used to model the diameter distribution, the most common including the two-parameter Weibull distribution, three-parameter Weibull distribution \citep{Bailey1973quantifying}, Johnson's SB (JSB) distribution \citep{fonseca2009describing}, and finite mixture distributions \citep{zasada2005finite}. Forest biometricians have developed a suite of methods for estimating the parameters of these models, including simple methods such as the method of moments and method of percentiles \citep{wang1995improved}, maximum likelihood techniques \citep{gove1989maximum}, and Bayesian methods \citep{green1994bayesian}. Another important tree and stand characterization used by foresters is the diameter to height relationship. Because diameter is an inexpensive field measurement relative to acquiring tree height, the forestry literature is rich with models to predict height based on diameter \citep{temesgen2014modelling}. Because of the extensive suite of techniques in the forestry literature to model diameter distributions and height-diameter relationships, there is a need for freely available software that facilitates streamlined model parameter estimation and comparison. To this end, we develop an open-source \proglang{R} package called \pkg{ForestFit} that provides a limited set of functions to fit a large variety of popular diameter distributions and height-diameter models. Importantly, the functions are designed to simplify comparison among models and parameter estimation techniques using a variety of model fit criteria, without requiring an in-depth knowledge of the statistical techniques used in the estimation process. 

In this paper, we describe the functionality of the \pkg{ForestFit} package and illustrate its features using the analysis of mixed ponderosa pine (\textit{Pinus ponderosa}) forest plots located in the Blue Mountains near Burns, Oregon, USA \citep{Kerns}. 


\section{\pkg{ForestFit} components}\label{section2}

In this section we provide an overview of the features available in the \pkg{ForestFit} package along with relevant citations that provide a more rigorous statistical treatment.

\subsection{Estimating parameters of the Weibull distribution}

The Weibull distribution is perhaps the most conspicuous distribution used for modeling diameter distributions due to its ability to represent common shapes in diameter distribution data \citep{Bailey1973quantifying,gorgoso2007modelling,gorgoso2014use,merganivc2006characterisation,pretsch2009forest,stankova2010modeling}. \pkg{ForestFit} offers several methods for estimating the parameters of two- and three-parameter Weibull distributions, which are listed in Table~\ref{tabWeibull} along with relevant sources for further details regarding each estimation method. 

\begin{table}[width=.9\linewidth,cols=4,pos=h]
\caption{Parameter estimation methods for two- and three- parameter Weibull distributions.} 
\centering 
\begin{tabular}{ccc} 
\toprule
Parameters & Method & Citation\\
\midrule
2 & Generalized regression type 1 & \cite{kantar2015generalized} \\
2 & Generalized regression type 2 & \cite{kantar2015generalized} \\
2 & L-moment & \cite{hosking1990moments} \\
2 & Maximum likelihood (ML) & \cite{gove1989maximum}\\
2 & Logarithmic moments & \cite{teimouri2013comparison}\\
2 & Method of moments & \cite{Bailey1973quantifying} \\
2 & Percentiles & \cite{wang1995improved} \\
2 & Rank correlation & \cite{teimouri2012simple} \\
2 & Least squares & \cite{kantar2015generalized} \\
2 & U-statistics & \cite{ustat} \\
2 & weighted ML & \cite{cousineau2009nearly} \\
2 & weighted least squares & \cite{zhang2008weighted} \\ 
3 & ML & \cite{green1994bayesian} \\
3 & Modified moments & \cite{cohen1982modified} \\
3 & Modified ML & \cite{cohen1982modified} \\
3 & Moments & \cite{cran1988moment} \\
3 & Maximum product spacing & \cite{teimouri2013comparison} \\
3 & weighted ML & \cite{cousineau2009nearly} \\
\bottomrule
\end{tabular} 
\label{tabWeibull}
\end{table}

\subsection{Bayesian analysis} 

Computational improvements over the last 30 years have led to the increase in popularity of Bayesian techniques for estimating model parameters for diameter distributions \citep{green1994bayesian}. In \pkg{ForestFit} we provide functionality for estimating the parameters of the three-parameter Weibull and four-parameter JSB distributions using the Bayesian paradigm. The JSB distribution is becoming increasingly popular for modeling diameter distributions in forestry \cite{fonseca2009describing, george2011estimation, zhang2003comparison} and is commonly applied in other fields, e.g., hydrology \cite{cugerone2015johnson} and climatology \cite{Lu2008JSB}.

With the exception of the location parameter, \pkg{ForestFit} uses the same algorithm as \cite{green1994bayesian} for estimating the shape and scale parameters of the three-parameter Weibull distribution.

\subsection{Modeling diameter distributions using grouped data}

Data collection protocols in forestry often lead to diameter observations that are recorded in groups or classes rather than as individual tree measurements. More specifically, suppose a data set has been classified into $m$ separate groups with lower bounds $r_0,r_1,\dots,r_m$ where $r_0=\min\bigl\{x_1,\dots,x_n\bigr\}$ and $r_m=\max\bigl\{x_1,\dots,x_n\bigr\}$, with $x_i$ being the individual diameter for tree $i$ out of a total $n$ trees in the sample. We define $f_i$, for $i=1,\dots,m$, as the frequency or number of trees in the $i$-th group. A schematic of grouped data is provided in Table \ref{tab1}.

\begin{table}[width=.9\linewidth,cols=4,pos=h]
\caption{A schematic of grouped data.} 
\centering 
\begin{tabular}{ccc} 
\midrule
Class number& Class& Frequency\\
\midrule
1&         $\bigl(r_{0}-r_{1}\bigr]$&$f_1$\\
2&         $(r_{1}-r_{2}\bigr]$&$f_{2}$\\
3&         $\bigl(r_{2}-r_{3}]$&$f_{3}$\\
\vdots& \vdots& \vdots\\
$m-1$& $\bigl(r_{m-2}-r_{m-1}\bigr]$&$f_{m-1}$\\
$m$   & $\bigl(r_{m-1}-r_{m}\bigr]$&$f_{m}$\\
\bottomrule
\end{tabular} 
\label{tab1}
\end{table}

\pkg{ForestFit} provides estimation of diameter distribution parameters for the three-parameter Birnbaum-Saunders (BS) distribution, the three-parameter generalized exponential (GE) distribution, and the three-parameter Weibull distribution using grouped data through the use of the vectors $r=(r_0,r_1,\dots,r_m)^{'}$ and $f=(f_1,f_2,\dots,f_m)^{'}$.

\subsection{Modeling diameter distributions using finite mixture models}\label{mixturepdf}

Certain forest stands (e.g., uneven aged or disturbed stands) exhibit multimodal diameter distributions that are not adequately represented by a single probability distribution. In such settings, forest biometricians and statisticians often use finite mixture distributions to characterize complex and multimodal diameter distributions \cite{zhang2001finite, liu2002finite, zasada2005finite, zhang2006fitting, liu2014modeling}. The probability density function (pdf) of a $\text{K}$-component mixture model has the form  

\begin{equation}\label{mixpdf} 
g(x|\Theta)=\sum_{k=1}^{\text{K}}\omega_k f\bigl(x|\boldsymbol{\theta}_k\bigr), 
\end{equation} 

where $\Theta=\bigl(\boldsymbol{\omega}^{'} ,\boldsymbol{\theta}_{1}^{'},\dots, \boldsymbol{\theta}_{k}^{'}\bigr)^{'}$ in which $\boldsymbol{\theta}_{k}=(\alpha_k,\beta_k)^{'}$ is the parameter vector of the $k$-th component with pdf $f(\cdot|\boldsymbol{\theta}_k)$ and $\boldsymbol{\omega}=(\omega_1,\dots,\omega_{\text{K}})^{'}$ is the vector of weight parameters. The weight parameters $\omega_k$s are non-negative and sum to one, i.e., $\sum_{k=1}^{\text{K}} \omega_k=1$. \pkg{ForestFit} enables $f(\cdot|\boldsymbol{\theta}_k)$ to take the following forms for individual (i.e., ungrouped) data: Burr XII, Chen, Fisher, Fr\'{e}chet, gamma, Gompertz, log-logistic, log-normal, Lomax, skew-normal, and two-parameter Weibull (pdfs are given in Appendix~\ref{apa}). For grouped data, the finite mixture distribution can be composed of gamma, log-normal, skew-normal, and two-parameter Weibull distributions. 

Finite mixture distributions are increasing in popularity in all areas of environmental and ecological statistics, see, e.g., \cite{hooten}. To facilitate testing and simulation using the common mixture models implemented in \pkg{ForestFit}, the package includes functions to compute the probability density function, cumulative density function, quantile function, and simulate random variables from the finite mixture models composed of the aforementioned probability distributions. Further, \pkg{ForestFit} provides specific functions to compute the probability density function, cumulative density function, quantile function, and simulate random variables from the gamma shape mixture model that has received increased attention in forestry for characterizing diameter distributions \citep{venturini2008}. 

\subsection{Estimating the parameters of height-diameter models} 

In addition to modeling diameter distributions, characterizing the height-diameter relationship is a common task in forestry \cite{temesgen2007regional}. A variety of mathematical relationships have been proposed to model this relationship, the most common of which are implemented in \pkg{ForestFit} and listed with associated references in Table~\ref{tab2}.

\begin{table}[width=.9\linewidth,cols=4,pos=h]
\caption{Common mathematical relationships used to depict the height-diameter relationship in a forest stand. $H$ is the height of a tree, $D$ is the diameter of the tree, and $\beta_1, \beta_2, \beta_3$ are estimated parameters.} 
\centering 
\begin{tabular}{lcl} 
\midrule
Model & Reference& Formula\\
\midrule
 Chapman-Richards & \cite{richards1959flexible} & ${H=1.3+\beta_1+\frac{\beta_2}{D+\beta_3}}$ \\
 Gompertz & \cite{winsor1932gompertz} & ${H=1.3+\beta_1 e^{-\beta_2e^{-\beta_3 D}}}$ \\
 Hossfeld IV & \cite{zeide1993analysis} & ${H=1.3+\frac{\beta_1}{1+\frac{1}{\beta_2 D^{\beta_3}}}}$  \\
 Korf & \cite{flewelling1994considerations} & ${H=1.3+\beta_1 e^{-\beta_2D^{-\beta_3}}}$ \\
 logistic & \cite{pearl1920rate} & ${H=1.3+\frac{\beta_1}{1+\beta_2e^{-\beta_3D}}}$ \\
 Prodan & \cite{prodan1968spatial} & ${H=1.3+\frac{D^2}{\beta_1 D^2+\beta_2 D+\beta_3}}$ \\
 Ratkowsky & \cite{ratkowsky1990handbook} & ${H=1.3+\beta_1 e^{-\frac{\beta_2}{D+\beta_3}}}$ \\
 Sibbesen & \cite{huang1992comparison} & ${H=1.3+\beta_1 D^{\beta_2 D^{-\beta_3}}}$ \\
 Weibull & \cite{yang1978potential} & ${H=1.3+\beta_1\Bigl(1-e^{-\beta_2 D^{\beta_3}}\Bigr)}$ \\
\bottomrule
\end{tabular} 
\label{tab2}
\end{table}
\section{Illustrations}

Here we illustrate key features available in \pkg{ForestFit}. All models and estimation methods assume the data points (e.g., diameter of an individual tree) are independent and identically distributed. For illustration, we use a set of 0.08 ha plots comprising mixed ponderosa pine and western Juniper (\textit{Juniperus occidentalis}) located in the Malheur National Forest on the southern end of the Blue Mountains near Burns, Oregon, USA \citep{Kerns}. These data are freely available (\url{https://www.fs.usda.gov/rds/archive/catalog/RDS-2017-0041}) under the condition that users cite the reference \citep{Kerns}. Further, we include these data in \pkg{ForestFit} so the data can easily be obtained using the command \texttt{data(DBH)}. For illustrative purposes, we only use the diameter and height measurements at 1.3 m height of all live trees in plots 72, 57, 55, and 51, and refer to these values as \texttt{d72, d57, d55}, and \texttt{d51}, respectively. The heights of trees in plot 55 are referred to as \texttt{h55}. Below we provide the commands for loading the \texttt{DBH} data set and obtaining the data from plot 55.

\begin{knitrout}
\definecolor{shadecolor}{rgb}{0.969, 0.969, 0.969}\color{fgcolor}\begin{kframe}
\begin{alltt}
\hlkwd{library}\hlstd{(ForestFit)}
\hlkwd{data}\hlstd{(DBH)}
\hlstd{d} \hlkwb{<-} \hlstd{DBH[DBH[,} \hlnum{1}\hlstd{]} \hlopt{==} \hlnum{55}\hlstd{,} \hlnum{10}\hlstd{]}
\hlstd{h} \hlkwb{<-} \hlstd{DBH[DBH[,} \hlnum{1}\hlstd{]} \hlopt{==} \hlnum{55}\hlstd{,} \hlnum{11}\hlstd{]}
\hlstd{d55} \hlkwb{<-} \hlstd{d[}\hlopt{!}\hlkwd{is.na}\hlstd{(d)]}
\hlstd{h55} \hlkwb{<-} \hlstd{h[}\hlopt{!}\hlkwd{is.na}\hlstd{(h)]}
\end{alltt}
\end{kframe}
\end{knitrout}

\subsection{Estimating parameters of the Weibull distribution}

The function \code{fitWeibull} estimates parameters of the two- and three-parameter Weibull distribution and takes the form

\begin{knitrout}
\definecolor{shadecolor}{rgb}{0.969, 0.969, 0.969}\color{fgcolor}\begin{kframe}
\begin{alltt}
\hlkwd{fitWeibull}\hlstd{(data, location, method, starts)}
\end{alltt}
\end{kframe}
\end{knitrout}

where \code{data} is a vector of data observations, \code{location} indicates whether to use a three-parameter Weibull distribution (\code{TRUE}) or a two-parameter Weibull distribution (\code{FALSE}), \code{method} denotes the estimation method to use, and \code{starts} is a vector of starting values for the parameters to be estimated. Available estimation methods and their code in \pkg{ForestFit} are shown in Table~\ref{tabMethods}.

\begin{table}[width=.9\linewidth,cols=4,pos=h]
\caption{Parameter estimation methods for the Weibull distribution and associated code used in the \texttt{method} argument of the \texttt{fitWeibull} function. MM refers to modified moment, ML refers to maximum likelihood, and MML refers to modified maximum likelihood.} 
\centering 
\begin{tabular}{cc} 
\toprule
Method & Code\\
\midrule
Generalized regression type 1 & \verb+greg1+ \\
Generalized regression type & \verb+greg2+ \\
L-moment & \verb+lm+ \\
ML (2 parameters) & \verb+ml+ \\
Logarithmic moments & \verb+mlm+ \\
Method of moments & \verb+moment+ \\
Percentiles & \verb+pm+ \\
Rank correlation & \verb+rank+ \\
Least squares & \verb+reg+ \\
U-statistics & \verb+ustat+ \\
Weighted ML & \verb+wml+ \\
Weighted least squares & \verb+wreg+ \\
ML (3 parameters) & \verb+mle+ \\
MM Type 1 & \verb+mm1+ \\
MM Type 2 & \verb+mm2+ \\
MM Type 3 & \verb+mm3+ \\
MML type 1 & \verb+mml1+ \\
MML type 2 & \verb+mml2+ \\
ML type 3 & \verb+mml3+ \\
MML type 4 & \verb+mml4+ \\
Maximum product spacing & \verb+mps+ \\
T-L moment & \verb+tlm+ \\
\bottomrule
\end{tabular} 
\label{tabMethods}
\end{table}

The code below calls \code{fitWeibull} to fit the three parameter Weibull distribution using maximum product spacing (\code{mps}) to the plot 72 data, with the function output printed below.

\begin{knitrout}
\definecolor{shadecolor}{rgb}{0.969, 0.969, 0.969}\color{fgcolor}\begin{kframe}
\begin{alltt}
\hlstd{starts} \hlkwb{<-} \hlkwd{c}\hlstd{(}\hlnum{2}\hlstd{,}\hlnum{2}\hlstd{,}\hlnum{3}\hlstd{)}
\hlkwd{fitWeibull}\hlstd{(}\hlkwc{mydata} \hlstd{= d72,} \hlkwc{location} \hlstd{=} \hlnum{TRUE}\hlstd{,}
           \hlkwc{method} \hlstd{=} \hlstr{"mps"}\hlstd{,} \hlkwc{starts} \hlstd{= starts)}
\end{alltt}
\begin{verbatim}
$estimate
        alpha     beta       mu
[1,] 1.297011 18.57024 12.41853

$measures
          AIC     CAIC      BIC     HQIC
[1,] 238.7653 239.6542 243.0672 240.1676
            AD       CVM        KS
[1,] 0.9055053 0.1239423 0.1441135
     log.likelihood
[1,]      -116.3826
\end{verbatim}
\end{kframe}
\end{knitrout}

The first element of the output list, \code{estimate}, contains the point estimates for the model parameters. The second element of the list, \code{measures}, contains a series of goodness-of-fit measures including Akaike Information Criterion (\code{AIC}), Consistent Akaike Information Criterion (\code{CAIC}), Bayesian Information Criterion (\code{BIC}), Hannan-Quinn information criterion (\code{HQIC}), Anderson-Darling (\code{AD}), Cram\'{e}r-von Misses (\code{CVM}), Kolmogorov-Smirnov (\code{KS}), and log-likelihood (\code{log.likelihood}). These measures allow for comparison between different models and estimation techniques to determine the best fitting model according to the desired criteria.

\subsection{Bayesian analysis}
Bayesian inference for diameter distributions modeled using the JSB and Weibull distributions is provided via the \code{fitbayesJSB} and \code{fitbayesWeibull} functions, respectively. The \code{fitbayesJSB} function has the form

\begin{knitrout}
\definecolor{shadecolor}{rgb}{0.969, 0.969, 0.969}\color{fgcolor}\begin{kframe}
\begin{alltt}
\hlkwd{fitbayesJSB}\hlstd{(data,} \hlkwc{n.burn} \hlstd{=} \hlnum{8000}\hlstd{,} \hlkwc{n.simul} \hlstd{=} \hlnum{10000}\hlstd{)}
\end{alltt}
\end{kframe}
\end{knitrout}

where \code{data} is a vector of observations, \code{n.burn} is an integer representing the number of burn-in iterations (i.e., the number of iterations the Gibbs sampler takes to reach convergence), and \code{n.simul} is the total number of Gibbs sampler iterations. By default, \code{fitbayesJSB} uses 10,000 total iterations and 8,000 burn-in iterations. \code{fitbayesWeibull} has the same arguments. Below we call both functions using the diameters of all live trees in plot 72.

\begin{knitrout}
\definecolor{shadecolor}{rgb}{0.969, 0.969, 0.969}\color{fgcolor}\begin{kframe}
\begin{alltt}
\hlkwd{fitbayesJSB}\hlstd{(d72,} \hlkwc{n.burn}\hlstd{=}\hlnum{8000}\hlstd{,} \hlkwc{n.simul}\hlstd{=}\hlnum{10000}\hlstd{)}
\end{alltt}
\end{kframe}
\end{knitrout}

\begin{knitrout}
\definecolor{shadecolor}{rgb}{0.969, 0.969, 0.969}\color{fgcolor}\begin{kframe}
\begin{verbatim}
$estimate
         delta     gamma   lambda       xi
[1,] 0.6897204 0.4077035 43.51213 11.68084

$measures
            AD        CVM        KS
[1,] 0.5059267 0.06804388 0.1235231
     log.likelihood
[1,]      -113.4759
\end{verbatim}
\end{kframe}
\end{knitrout}

\begin{knitrout}
\definecolor{shadecolor}{rgb}{0.969, 0.969, 0.969}\color{fgcolor}\begin{kframe}
\begin{alltt}
\hlkwd{fitbayesWeibull}\hlstd{(d72,} \hlkwc{n.burn}\hlstd{=}\hlnum{8000}\hlstd{,} \hlkwc{n.simul}\hlstd{=}\hlnum{10000}\hlstd{)}
\end{alltt}
\end{kframe}
\end{knitrout}

\begin{knitrout}
\definecolor{shadecolor}{rgb}{0.969, 0.969, 0.969}\color{fgcolor}\begin{kframe}
\begin{verbatim}
$estimate
        alpha     beta       mu
[1,] 1.632042 21.18533 10.28903

$measures
            AD       CVM        KS
[1,] 0.7636954 0.1028456 0.1378961
     log.likelihood
[1,]      -116.7509
\end{verbatim}
\end{kframe}
\end{knitrout}

\begin{knitrout}
\definecolor{shadecolor}{rgb}{0.969, 0.969, 0.969}\color{fgcolor}\begin{figure}
\includegraphics[width=\maxwidth]{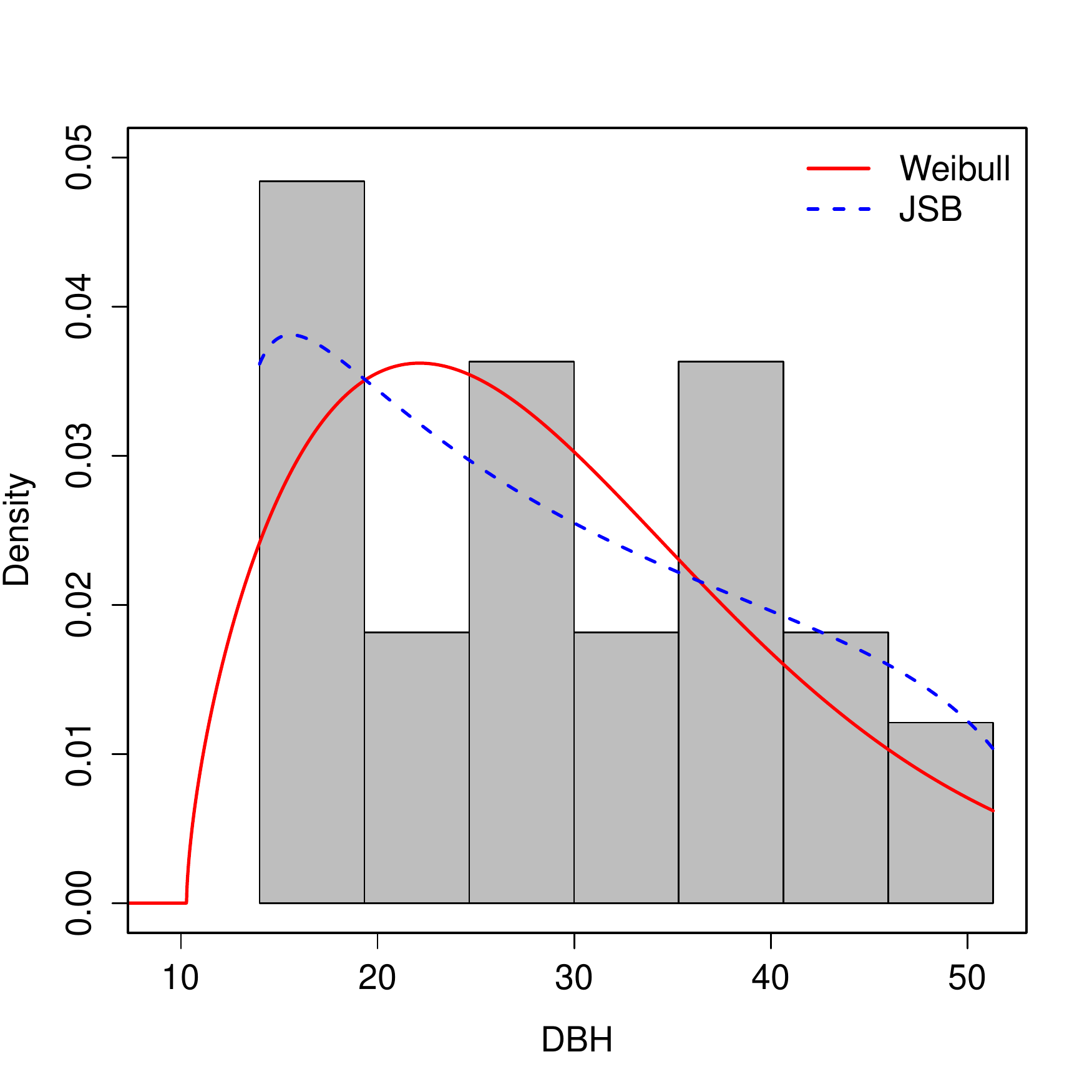} \caption[Diameter distribution of trees in plot 72]{Diameter distribution of trees in plot 72. Fitted JSB and Weibull pdfs estimated using the Bayesian paradigm are superimposed.}\label{fig:bayesOut}
\end{figure}

\end{knitrout}

We see that for plot 72, all goodness-of-fit measures indicate the JSB model is a better fit for the data (Figure \ref{fig:bayesOut}).

\subsection{Modeling diameter distributions using grouped data}\label{groupedCode}

Suppose diameter observations are given in \code{m} separate groups as shown in Table~\ref{tab1}. \pkg{ForestFit} provides the function \code{fitgrouped} for estimating the parameters of the three-parameter BS and Weibull distributions using grouped data. This function takes the form

\begin{knitrout}
\definecolor{shadecolor}{rgb}{0.969, 0.969, 0.969}\color{fgcolor}\begin{kframe}
\begin{alltt}
\hlkwd{fitgrouped}\hlstd{(r, f, family, method1, starts, method2)}
\end{alltt}
\end{kframe}
\end{knitrout}

where \code{r} is a length $m+1$ vector of group boundaries as shown in Table~\ref{tab1}, with the first element of \code{r} being the lower bound of the first group and all other \code{m} values representing the upper bound of the \code{m} groups. \code{f} is a vector of length $m$ of the group frequencies. \code{family} represents the distribution used in the model, taking values \code{`weibull'} or \code{`birnbaum-saunders'}. \code{method1} is a string denoting the method of estimation. Here we enable three methods of estimation: approximated maximum likelihood (\code{`aml'}), EM algorithm (\code{`em'}), and maximum likelihood (\code{`ml'}). \code{starts} is a numeric vector containing the starting values for the shape, scale, and location parameters, respectively. Lastly, \code{method2} indicates the optimization method of the log-likelihood, taking values \code{`BFGS'}, \code{`CG'}, \code{`L-BFGS-B'}, \code{`Nelder-Mean'}, and \code{`SANN'}. If using the EM algorithm, the \code{stats} and \code{method2} arguments do not need to be specified. Below we manually group the diameter observations from plot 57 into six groups, and use the function \code{fitgrouped} on this grouped data set.

\begin{knitrout}
\definecolor{shadecolor}{rgb}{0.969, 0.969, 0.969}\color{fgcolor}\begin{kframe}
\begin{alltt}
\hlstd{m} \hlkwb{<-} \hlnum{6}
\hlstd{r} \hlkwb{<-} \hlkwd{seq}\hlstd{(}\hlkwd{min}\hlstd{(d57),} \hlkwd{max}\hlstd{(d57),} \hlkwc{length}\hlstd{=m}\hlopt{+}\hlnum{1}\hlstd{)}
\hlstd{D} \hlkwb{<-} \hlkwd{data.frame}\hlstd{(}\hlkwd{table}\hlstd{(}\hlkwd{cut}\hlstd{(d57, r)))}
\hlkwd{fitgrouped}\hlstd{(}\hlkwc{r} \hlstd{= r,} \hlkwc{f} \hlstd{= D}\hlopt{$}\hlstd{Freq,}
           \hlkwc{family} \hlstd{=} \hlstr{"birnbaum-saunders"}\hlstd{,}
           \hlkwc{method1} \hlstd{=} \hlstr{"em"}\hlstd{)}
\end{alltt}
\end{kframe}
\end{knitrout}

\begin{knitrout}
\definecolor{shadecolor}{rgb}{0.969, 0.969, 0.969}\color{fgcolor}\begin{kframe}
\begin{verbatim}
$estimate
         alpha     beta       mu
[1,] 0.6234071 8.660411 8.453387

$measures
          AIC     CAIC      BIC    HQIC
[1,] 24.81261 25.03483 28.89871 26.4006
           AD Chi-square      CVM         KS
[1,] 8.401981   1.595622 4.393473 0.05225002
     log.likelihood
[1,]       -10.4063
\end{verbatim}
\end{kframe}
\end{knitrout}

\begin{knitrout}
\definecolor{shadecolor}{rgb}{0.969, 0.969, 0.969}\color{fgcolor}\begin{figure}
\includegraphics[width=\maxwidth]{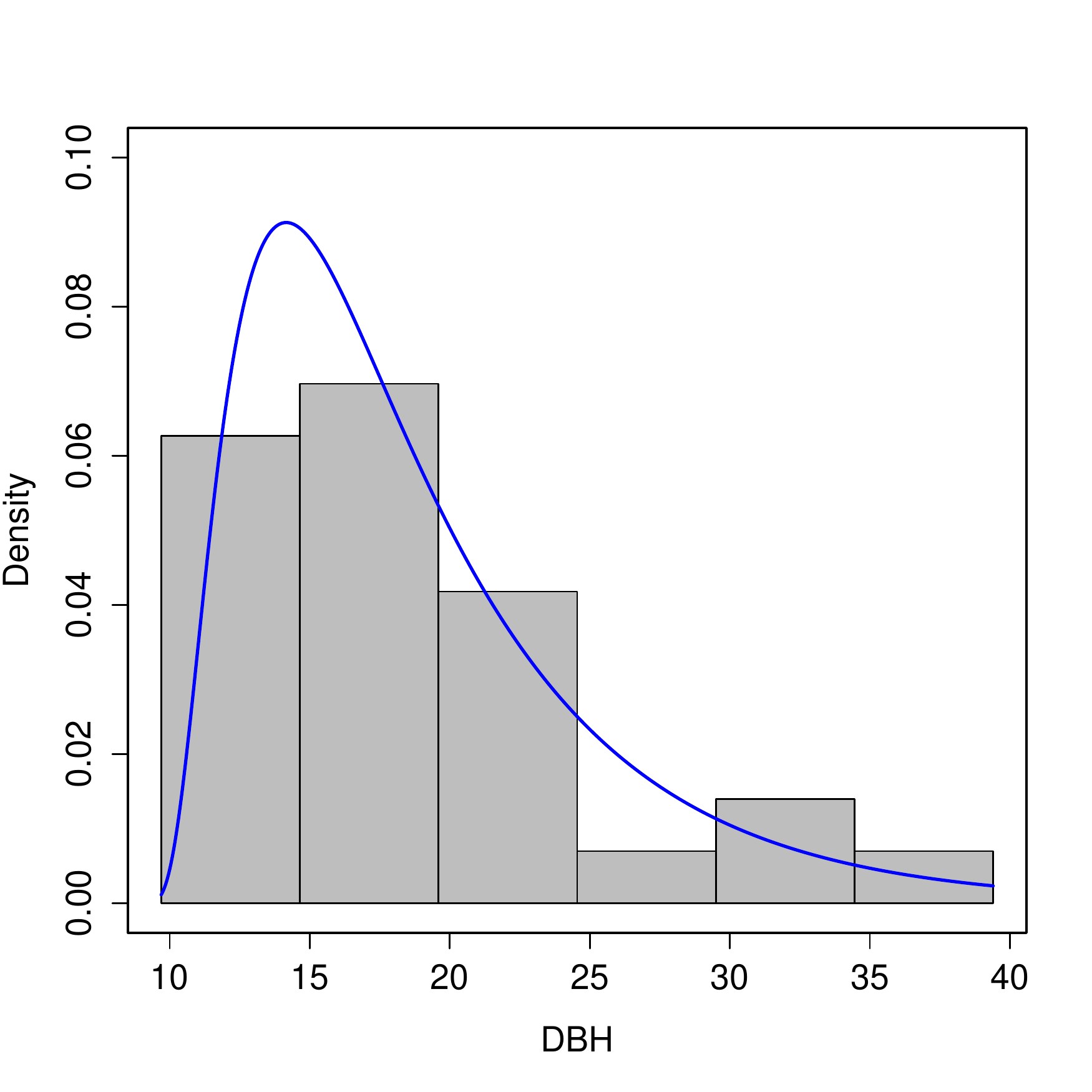} \caption[Diameter distribution of trees in plot 57 classified into six groups]{Diameter distribution of trees in plot 57 classified into six groups. The estimated pdf of the BS distribution is superimposed.}\label{fig:grouped}
\end{figure}

\end{knitrout}

Figure~\ref{fig:grouped} displays the estimated BS pdf to plot 57's diameter distribution data.

\subsection{Modeling diameter distributions of grouped data using finite mixture models}\label{mixturegrouped}

The \code{fitmixturegrouped} function provides an interface for fitting a wide-range of finite mixture distributions to grouped data and takes the form

\begin{knitrout}
\definecolor{shadecolor}{rgb}{0.969, 0.969, 0.969}\color{fgcolor}\begin{kframe}
\begin{alltt}
\hlkwd{fitmixturegrouped}\hlstd{(family, r, f, K,}
                  \hlkwc{initial} \hlstd{=} \hlnum{FALSE}\hlstd{, starts)}
\end{alltt}
\end{kframe}
\end{knitrout}

where \code{r} and \code{f} are vectors of the class boundaries and frequencies as previously described in Section \ref{groupedCode}. \code{K} is a single numeric value denoting the number of component pdfs to include in the finite mixture model. All parameters in finite mixture models are estimated using the EM algorithm. The \code{family} argument is a character string denoting the distribution used in the mixture model. Currently available distributions are the gamma (\code{`gamma'}), log-normal (\code{`log-normal'}), skew-normal (\code{`skew-normal'}), and Weibull (\code{`weibull'}) families.  The \code{initial} argument denotes whether or not the user wants to specify starting values for the EM algorithm. By default the argument is set to \code{FALSE} and thus the user does not need to specify a value for the \code{starts} argument. If \code{initial=TRUE}, the user must specify a sequence of initial values taking the form \code{starts} = $(\bm{\omega}, \bm{\alpha}, \bm{\beta})$, where the three vectors are of length K with elements $\omega_i$ being the weight of the $i$-th component, and $\alpha_i$ and $\beta_i$ are the associated parameters of the $i$-th component. If \code{family = `skew-normal'}, the user must specify starting values for the weights and three parameters for each component distribution (see Appendix~\ref{apa}). In the code that follows, we illustrate the \code{fitmixturegrouped} function by fitting a two-component skew-normal mixture model to plot 51's grouped data.

\begin{knitrout}
\definecolor{shadecolor}{rgb}{0.969, 0.969, 0.969}\color{fgcolor}\begin{kframe}
\begin{alltt}
\hlstd{m} \hlkwb{<-} \hlnum{8}\hlstd{; K} \hlkwb{<-} \hlnum{2}
\hlstd{r} \hlkwb{<-} \hlkwd{seq}\hlstd{(}\hlkwd{min}\hlstd{(d51),}\hlkwd{max}\hlstd{(d51),}\hlkwc{length}\hlstd{=m}\hlopt{+}\hlnum{1}\hlstd{)}
\hlstd{D} \hlkwb{<-} \hlkwd{data.frame}\hlstd{(}\hlkwd{table}\hlstd{(}\hlkwd{cut}\hlstd{(d51,r)))}
\hlstd{f} \hlkwb{<-} \hlstd{D}\hlopt{$}\hlstd{Freq}
\hlstd{omega} \hlkwb{<-} \hlkwd{c}\hlstd{(}\hlnum{0.5}\hlstd{,}\hlnum{0.5}\hlstd{); alpha} \hlkwb{<-} \hlkwd{c}\hlstd{(}\hlnum{10}\hlstd{,}\hlnum{40}\hlstd{)}
\hlstd{beta} \hlkwb{<-} \hlkwd{c}\hlstd{(}\hlnum{2}\hlstd{,}\hlnum{2}\hlstd{); lambda} \hlkwb{<-} \hlkwd{c}\hlstd{(}\hlnum{2}\hlstd{,}\hlopt{-}\hlnum{2}\hlstd{)}
\hlkwd{fitmixturegrouped}\hlstd{(}\hlkwc{family} \hlstd{=} \hlstr{"skew-normal"}\hlstd{,}
                  \hlkwc{r} \hlstd{= r,} \hlkwc{f} \hlstd{= f,}
                  \hlkwc{K} \hlstd{= K,} \hlkwc{initial} \hlstd{=} \hlnum{TRUE}\hlstd{,}
                  \hlkwc{starts} \hlstd{=} \hlkwd{c}\hlstd{(omega, alpha,}
                             \hlstd{beta, lambda))}
\end{alltt}
\end{kframe}
\end{knitrout}

\begin{knitrout}
\definecolor{shadecolor}{rgb}{0.969, 0.969, 0.969}\color{fgcolor}\begin{kframe}
\begin{verbatim}
$estimate
        weight    alpha     beta   lambda
[1,] 0.6296296 12.11372 1.504074 4.920060
[2,] 0.3703704 32.74539 8.368941 5.843043

$measures
          AIC    CAIC      BIC     HQIC
[1,] 27.82552 30.2603 41.74841 33.19503
           AD Chi-square      CVM         KS
[1,] 21.06323   1.060218 4.625018 0.03171701
     log.likelihood
[1,]      -6.912759
\end{verbatim}
\end{kframe}
\end{knitrout}

\begin{knitrout}
\definecolor{shadecolor}{rgb}{0.969, 0.969, 0.969}\color{fgcolor}\begin{figure}
\includegraphics[width=\maxwidth]{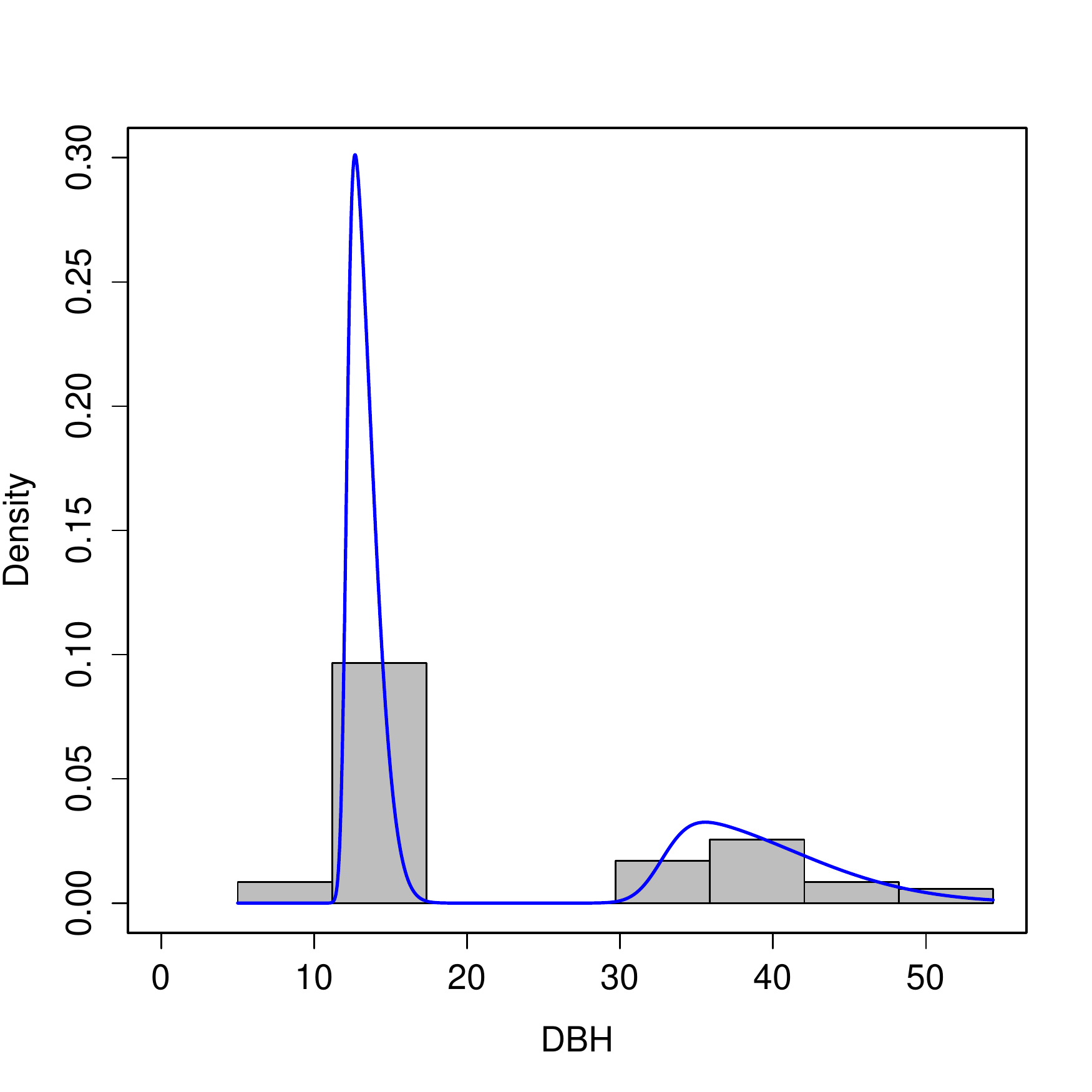} \caption[Histogram of grouped diameter distribution data in plot 51]{Histogram of grouped diameter distribution data in plot 51. Superimposed is the estimated pdf of the two-component skew-normal mixture model.}\label{fig:groupedMix}
\end{figure}

\end{knitrout}

Similar to previous functions using models with a single pdf, the output is a list where the first element (i.e., \code{estimate}) contains the estimated parameters of the first and second components of the model, and the second element (i.e., \code{measures}) contains the goodness-of-fit measures. Figure~\ref{fig:groupedMix} displays the estimated two-component skew-normal pdf.

\subsection{Modeling diameter distributions using finite mixture models fitted to ungrouped data}\label{mixturepdfGrouped}

As a companion to the grouped data \code{fitmixturegrouped} function described in Section~\ref{mixturegrouped}, the \code{fitmixture} function fits finite mixture models to individual diameter distribution data, and takes the form

\begin{knitrout}
\definecolor{shadecolor}{rgb}{0.969, 0.969, 0.969}\color{fgcolor}\begin{kframe}
\begin{alltt}
\hlkwd{fitmixture}\hlstd{(data, family, K,} \hlkwc{initial} \hlstd{=} \hlnum{FALSE}\hlstd{, starts)}
\end{alltt}
\end{kframe}
\end{knitrout}

where \code{data} is a vector of individual diameter measurements and \code{family} denotes the family of probability distributions to use. Available families of pdfs are the BS \\(\code{`birnbaum-saunders'}), Burr XI (\code{`burrxii'}), Chen (\code{`chen'}), Fisher (\code{`f'}), Fr\'{e}chet (\code{`Frechet'}), gamma (\code{`gamma'}), generalized exponential (\code{`ge'}), Gompertz (\code{`gompertz'}), log-normal (\code{`log-normal'}), log-logistic (\code{`log-logistic'}), Lomax (\code{`lomax'}), skew-normal (\code{`skew-normal'}), and Weibull (\code{`weibull'}) families. The remaining arguments are analogous to those used in the \code{fitmixturegrouped} function. Below, the \code{fitmixture} function is called to fit a two-component log-normal mixture model to plot 51's diameter data. 

\begin{knitrout}
\definecolor{shadecolor}{rgb}{0.969, 0.969, 0.969}\color{fgcolor}\begin{kframe}
\begin{alltt}
\hlkwd{fitmixture}\hlstd{(d51,} \hlstr{"log-normal"}\hlstd{,} \hlnum{2}\hlstd{)}
\end{alltt}
\end{kframe}
\end{knitrout}

\begin{knitrout}
\definecolor{shadecolor}{rgb}{0.969, 0.969, 0.969}\color{fgcolor}\begin{kframe}
\begin{verbatim}
$estimate
        weight    alpha      beta
[1,] 0.6522847 2.618974 0.2998769
[2,] 0.3477153 3.668380 0.1461719

$measures
          AIC     CAIC      BIC     HQIC
[1,] 419.7617 420.9381 429.9769 423.7317
           AD      CVM        KS
[1,] 5.790067 1.107157 0.3496867
     log.likelihood
[1,]      -204.8808

$cluster
 [1] 1 1 1 2 2 2 2 2 1 2 1 2 1 1 1 1 1 1 1 1
[21] 1 1 2 2 2 2 1 1 1 1 1 1 1 1 1 1 2 1 1 1
[41] 2 2 1 1 2 2 2 2 2 1 1 2 1 1 1 1 1
\end{verbatim}
\end{kframe}
\end{knitrout}

\begin{knitrout}
\definecolor{shadecolor}{rgb}{0.969, 0.969, 0.969}\color{fgcolor}\begin{figure}
\includegraphics[width=\maxwidth]{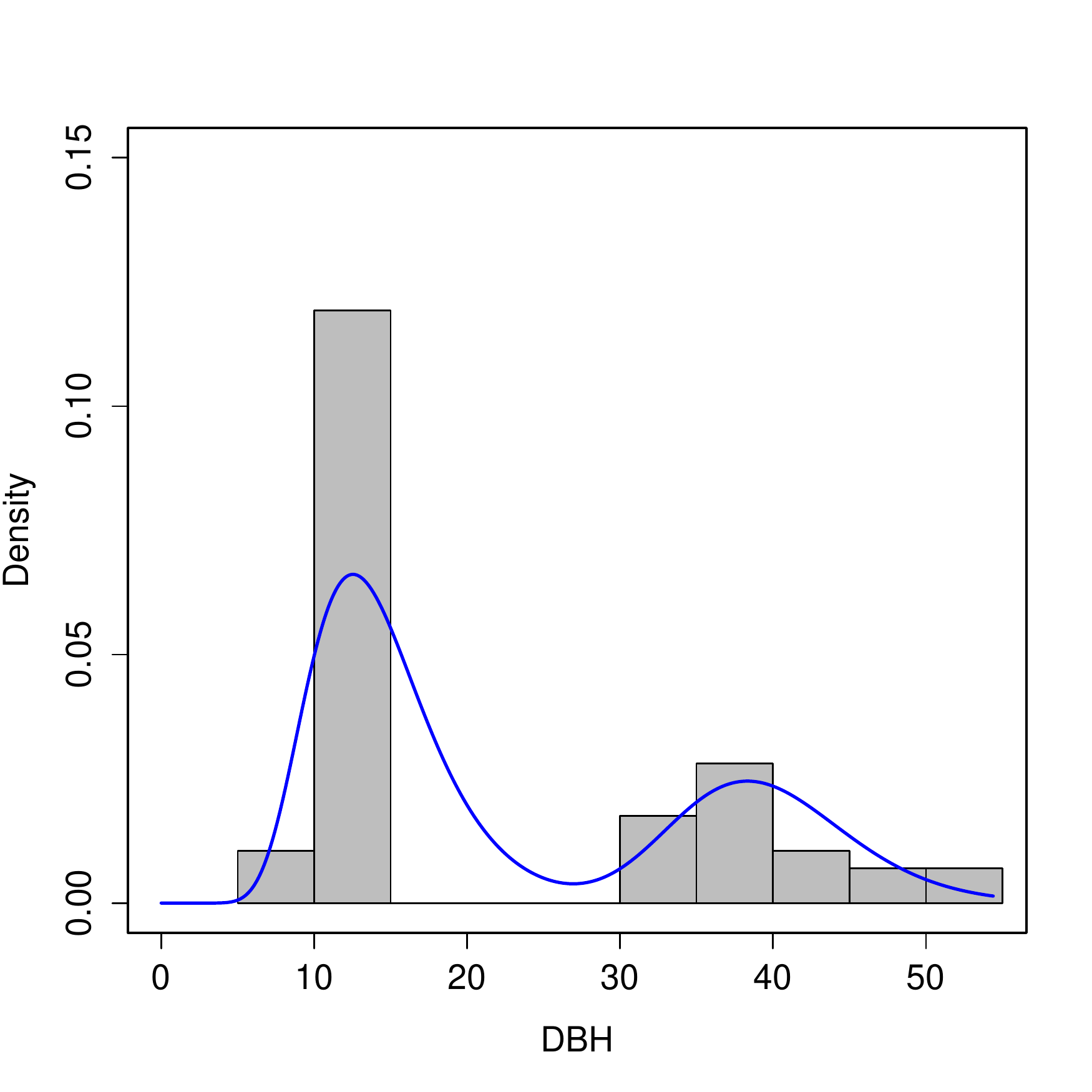} \caption[Diameter distribution of plot 51]{Diameter distribution of plot 51. Superimposed is the estimated pdf of the two-component log-normal mixture model.}\label{fig:ungroupedMixed}
\end{figure}

\end{knitrout}

Interpretation of the first and second components of the output list is analagous to the interpretation provided in Section~\ref{mixturegrouped}. The third list element (\code{cluster}) specifies the component of the mixture distribution from which each individual diameter measurement arises. Figure~\ref{fig:ungroupedMixed} displays the fitted two-component log-normal pdf to the diameter distribution data of plot 51. 

\subsection{Distribution functions for finite mixture models}\label{mixturepdfFuncs}

Analogous with \proglang{R}'s standard functions for working with common probability distributions, we provide commands for computing the density, distribution function, quantile function, and random generation from the finite mixture models used in \code{ForestFit}. The functions take the following forms

\begin{knitrout}
\definecolor{shadecolor}{rgb}{0.969, 0.969, 0.969}\color{fgcolor}\begin{kframe}
\begin{alltt}
\hlkwd{dmixture}\hlstd{(x, family, K, param)}
\hlkwd{pmixture}\hlstd{(x, family, K, param)}
\hlkwd{qmixture}\hlstd{(p, family, K, param)}
\hlkwd{rmixture}\hlstd{(n, family, K, param)}
\end{alltt}
\end{kframe}
\end{knitrout}

where \code{x} is a vector of observations, \code{n} is the number of realizations to be simulated from the mixture model, \code{p} is a numeric value that satisfies $0<p<1$, \code{family} is one of the families introduced in Section~\ref{mixturepdf}, \code{K} is the number of components, and \code{param} is the mixture model parameter vector that takes the same form as the \code{starts} argument of the \code{fitmixturegrouped} function described previously in Section~\ref{mixturegrouped}. As an illustration, the code below generates 500 realizations from a three-component BS mixture model and displays the resulting distribution in Figure~\ref{fig:figSim}:

\begin{knitrout}
\definecolor{shadecolor}{rgb}{0.969, 0.969, 0.969}\color{fgcolor}\begin{kframe}
\begin{alltt}
\hlstd{n} \hlkwb{<-} \hlnum{500}\hlstd{; K} \hlkwb{<-} \hlnum{3}
\hlstd{weight} \hlkwb{<-} \hlkwd{c}\hlstd{(}\hlnum{0.4}\hlstd{,} \hlnum{0.3}\hlstd{,} \hlnum{0.3}\hlstd{)}
\hlstd{alpha} \hlkwb{<-} \hlkwd{c}\hlstd{(}\hlnum{0.1}\hlstd{,} \hlnum{0.25}\hlstd{,} \hlnum{0.5}\hlstd{)}
\hlstd{beta} \hlkwb{<-} \hlkwd{c}\hlstd{(}\hlnum{8}\hlstd{,} \hlnum{5}\hlstd{,} \hlnum{2}\hlstd{)}
\hlstd{param} \hlkwb{<-} \hlkwd{c}\hlstd{(weight, alpha, beta)}
\hlstd{X} \hlkwb{<-} \hlkwd{rmixture}\hlstd{(n,} \hlstr{"birnbaum-saunders"}\hlstd{,}
              \hlstd{K, param)}
\hlstd{XX} \hlkwb{<-} \hlkwd{seq}\hlstd{(}\hlnum{0}\hlstd{,} \hlkwd{max}\hlstd{(X),} \hlnum{0.01}\hlstd{)}
\hlstd{pdf} \hlkwb{<-} \hlkwd{dmixture}\hlstd{(XX,} \hlstr{"birnbaum-saunders"}\hlstd{,}
                \hlstd{K, param)}
\hlkwd{hist}\hlstd{(X,} \hlkwc{freq}\hlstd{=}\hlnum{FALSE}\hlstd{,} \hlkwc{breaks}\hlstd{=}\hlnum{120}\hlstd{,} \hlkwc{col}\hlstd{=}\hlstr{'gray'}\hlstd{,}
     \hlkwc{las} \hlstd{=} \hlnum{1}\hlstd{,} \hlkwc{main} \hlstd{=} \hlstr{''}\hlstd{,} \hlkwc{xlim} \hlstd{=} \hlkwd{c}\hlstd{(}\hlnum{0}\hlstd{,} \hlnum{11}\hlstd{))}
\hlkwd{lines}\hlstd{(XX, pdf,} \hlkwc{col}\hlstd{=}\hlstr{'blue'}\hlstd{)}
\end{alltt}
\end{kframe}\begin{figure}
\includegraphics[width=\maxwidth]{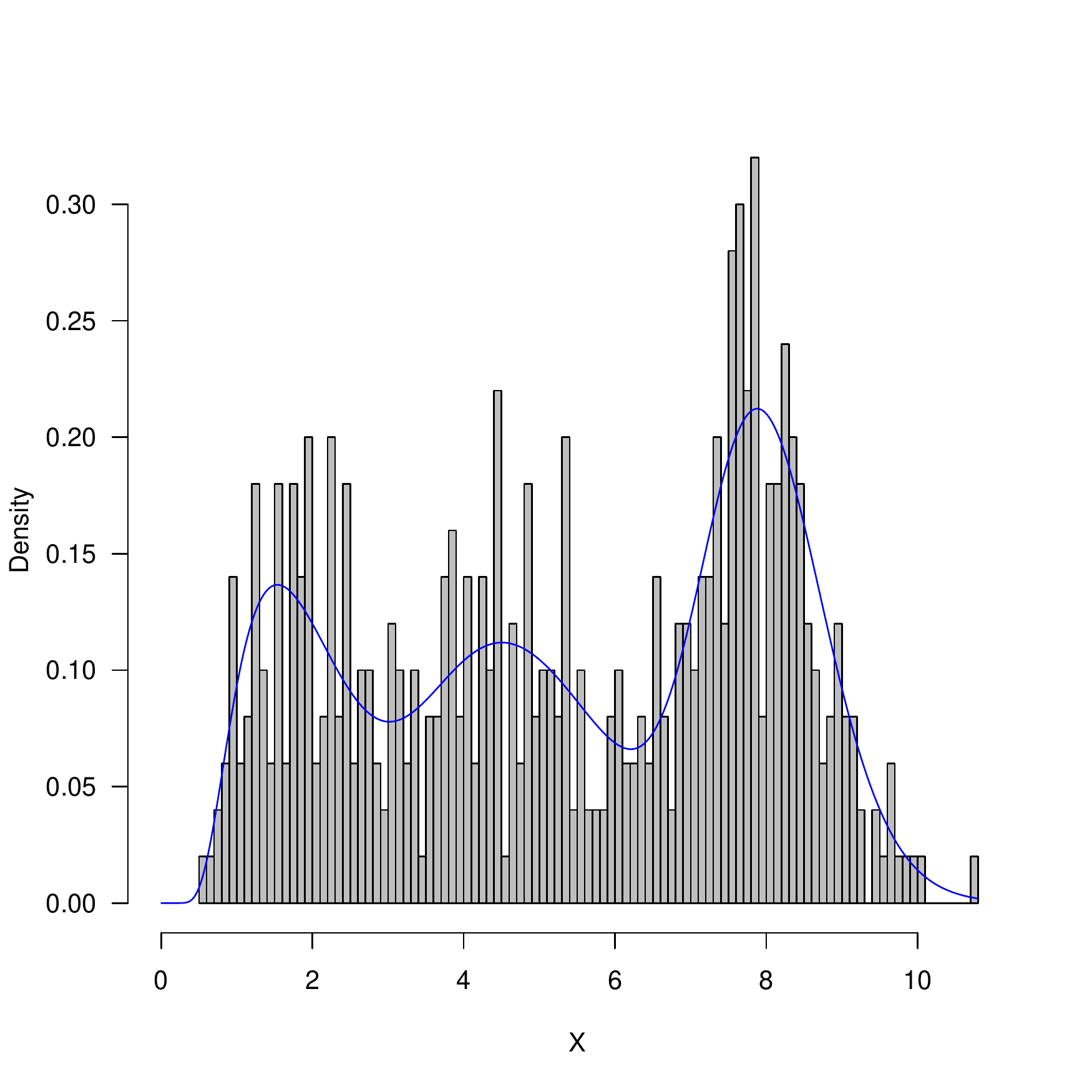} \caption[Histogram of 500 random realizations from a three-component BS mixture model]{Histogram of 500 random realizations from a three-component BS mixture model.}\label{fig:figSim}
\end{figure}

\end{knitrout}

\subsection{The gamma shape mixture (GSM) model}

The gamma shape mixture model (GSM) introduced by \citep{venturini2008} has received considerable attention in forestry for characterizing diameter distributions. Because of its popularity, we have implemented specific functions in \pkg{ForestFit} for estimating the parameters of the GSM distribution, computing the density function, computing the distribution function, and generating realizations for the GSM distribution. The function calls for the aforementioned tasks are

\begin{knitrout}
\definecolor{shadecolor}{rgb}{0.969, 0.969, 0.969}\color{fgcolor}\begin{kframe}
\begin{alltt}
\hlkwd{fitgsm}\hlstd{(data, K)}
\hlkwd{dgsm}\hlstd{(data, omega, beta,} \hlkwc{log} \hlstd{=} \hlnum{FALSE}\hlstd{)}
\hlkwd{pgsm}\hlstd{(data, omega, beta,} \hlkwc{log.p} \hlstd{=} \hlnum{FALSE}\hlstd{,} \hlkwc{lower.tail} \hlstd{=} \hlnum{TRUE}\hlstd{)}
\hlkwd{rgsm}\hlstd{(n, omega, beta)}
\end{alltt}
\end{kframe}
\end{knitrout}

where \code{data} is a vector of observations, \code{n} is the number of realizations simulated from the GSM model, \code{K} is the number of components, \code{omega} is a vector of mixing parameters, \code{beta} is the rate parameter, \code{log} indicates whether to compute the density function (\code{log = FALSE}) or the log-density function (\code{log = TRUE}), \code{log.p} indicates whether to compute the the distribution function (\code{log.p = FALSE}) or the log-transformed distribution function (\code{log.p = TRUE}), and \code{lower.tail} indicates whether to compute the upper tail (\code{lower.tail = FALSE}) or lower tail of the distribution (\code{lower.tail = TRUE}). As an illustration, we generate 500 realizations from a ten-component GSM model and display the resulting distribution in Figure~\ref{fig:gsmCurve}.

\begin{knitrout}
\definecolor{shadecolor}{rgb}{0.969, 0.969, 0.969}\color{fgcolor}\begin{kframe}
\begin{alltt}
\hlstd{n} \hlkwb{<-} \hlnum{500}
\hlstd{K} \hlkwb{<-} \hlnum{10}
\hlstd{omega} \hlkwb{<-} \hlkwd{rep}\hlstd{(}\hlnum{0.1}\hlstd{,} \hlnum{10}\hlstd{)}
\hlstd{beta} \hlkwb{<-} \hlnum{0.25}
\hlstd{X} \hlkwb{<-} \hlkwd{rgsm}\hlstd{(n, omega, beta)}
\hlstd{XX} \hlkwb{<-} \hlkwd{seq}\hlstd{(}\hlnum{0}\hlstd{,} \hlkwd{max}\hlstd{(X),} \hlnum{0.01}\hlstd{)}
\hlstd{out} \hlkwb{<-} \hlkwd{fitgsm}\hlstd{(X, K)}
\hlstd{pdf} \hlkwb{<-} \hlkwd{dgsm}\hlstd{(XX, out}\hlopt{$}\hlstd{omega, out}\hlopt{$}\hlstd{beta)}
\hlkwd{hist}\hlstd{(X,} \hlkwc{freq} \hlstd{=} \hlnum{FALSE}\hlstd{,} \hlkwc{breaks} \hlstd{=} \hlnum{30}\hlstd{,} \hlkwc{col} \hlstd{=} \hlstr{'gray'}\hlstd{,}
     \hlkwc{main} \hlstd{=} \hlstr{''}\hlstd{,} \hlkwc{xlim} \hlstd{=} \hlkwd{c}\hlstd{(}\hlnum{0}\hlstd{,} \hlnum{80}\hlstd{))}
\hlkwd{lines}\hlstd{(XX, pdf,} \hlkwc{col} \hlstd{=} \hlstr{'blue'}\hlstd{)}
\end{alltt}
\end{kframe}\begin{figure}
\includegraphics[width=\maxwidth]{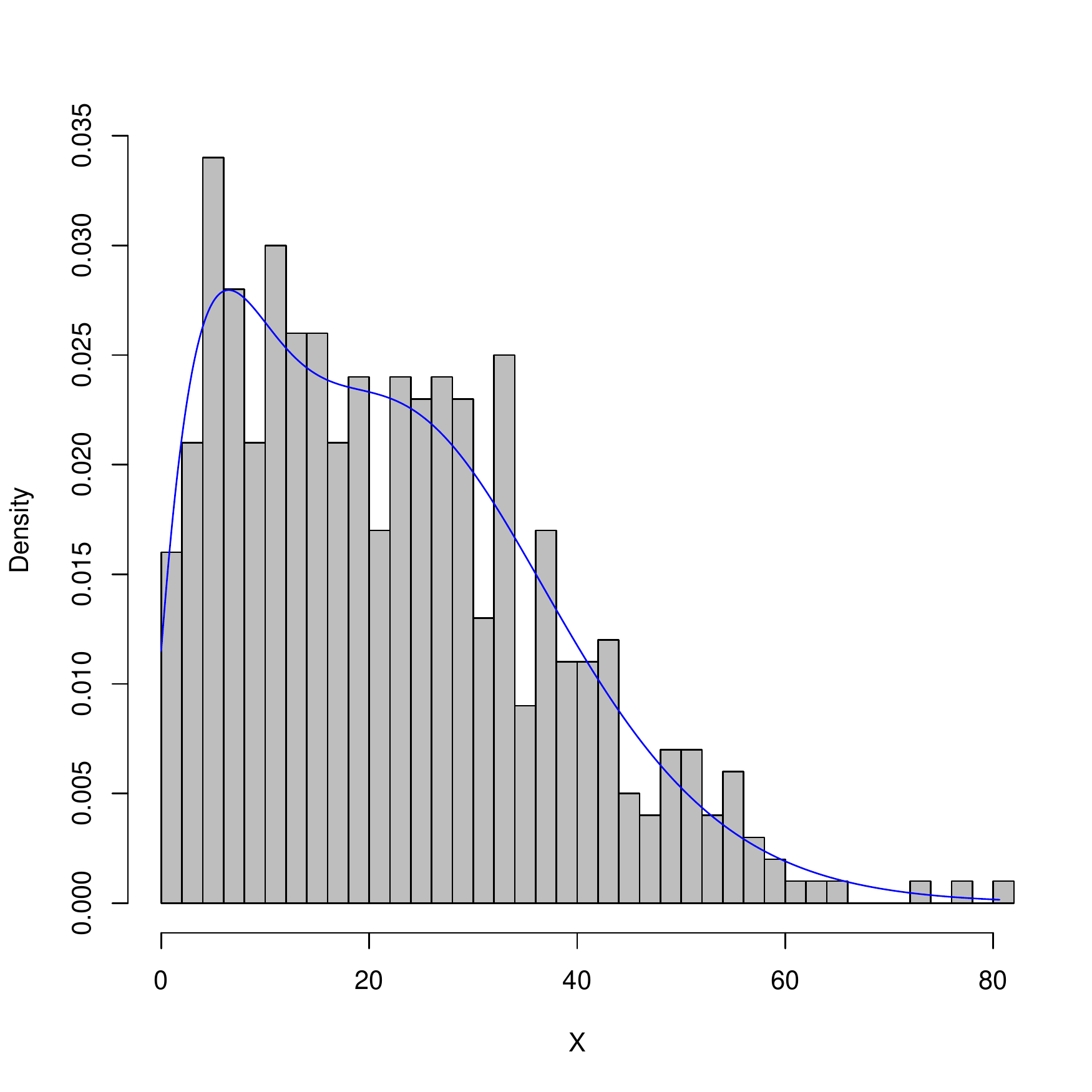} \caption[Histogram of 500 relaizations from a ten-component GSM model]{Histogram of 500 relaizations from a ten-component GSM model.}\label{fig:gsmCurve}
\end{figure}

\begin{kframe}\begin{alltt}
\hlstd{out}
\end{alltt}
\begin{verbatim}
$beta
[1] 0.2071086

$omega
 [1] 5.553682e-02 2.005843e-01 1.948963e-01
 [4] 9.702007e-03 2.543254e-02 1.802041e-01
 [7] 3.084612e-01 2.515575e-02 2.694728e-05
[10] 8.429742e-11

$measures
          AIC     CAIC      BIC     HQIC
[1,] 4061.953 4065.655 4184.176 4109.913
            AD        CVM        KS
[1,] 0.1346958 0.01655488 0.0167851
     log.likelihood
[1,]      -2001.976
\end{verbatim}
\end{kframe}
\end{knitrout}

The output list of the \texttt{fitgsm} function consists of three parts: 

\begin{enumerate}
\item \texttt{beta}: estimated rate parameter.
\item \texttt{omega}: estimated vector of mixing parameters.
\item \texttt{measures}: sequence of goodness-of-fit measures including \texttt{AIC, CAIC, HQIC, AD, CVM, KS} and \texttt{log.likelihood}. 
\end{enumerate}

\subsection{Modeling height-diameter relationships}\label{heightDiameter}

The general function for fitting a growth curve to paired observations of height and diameter is

\begin{knitrout}
\definecolor{shadecolor}{rgb}{0.969, 0.969, 0.969}\color{fgcolor}\begin{kframe}
\begin{alltt}
\hlkwd{fitgrowth}\hlstd{(h, d, model, starts)}
\end{alltt}
\end{kframe}
\end{knitrout}

where \code{h} and \code{d} are vectors of height and diameter observations, respectively. The \code{model} argument denotes the model to fit with options: Chapman-Richards (\code{`chapman-richards'}), Gompertz (\code{`gompertz'}), Hossfeld IV (\code{`hossfeldiv'}), Korf (\code{`korf'}), logistic (\code{`logistic'}), Prodan (\code{`prodan'}) , Ratkowsky (\code{`ratkowsky'}), Sibbesen (\code{`Sibbesen'}), and Weibull (\code{`weibull'}). Here we fit a Weibull growth curve to the height-diameter relationship of plot 55. 

\begin{knitrout}
\definecolor{shadecolor}{rgb}{0.969, 0.969, 0.969}\color{fgcolor}\begin{kframe}
\begin{alltt}
\hlstd{starts} \hlkwb{<-} \hlkwd{c}\hlstd{(}\hlnum{18}\hlstd{,} \hlnum{0.0005}\hlstd{,} \hlnum{2}\hlstd{)}
\hlkwd{fitgrowth}\hlstd{(h55, d55,} \hlstr{"weibull"}\hlstd{,} \hlkwc{starts}\hlstd{=starts)}
\end{alltt}
\end{kframe}\begin{figure}
\includegraphics[width=\maxwidth]{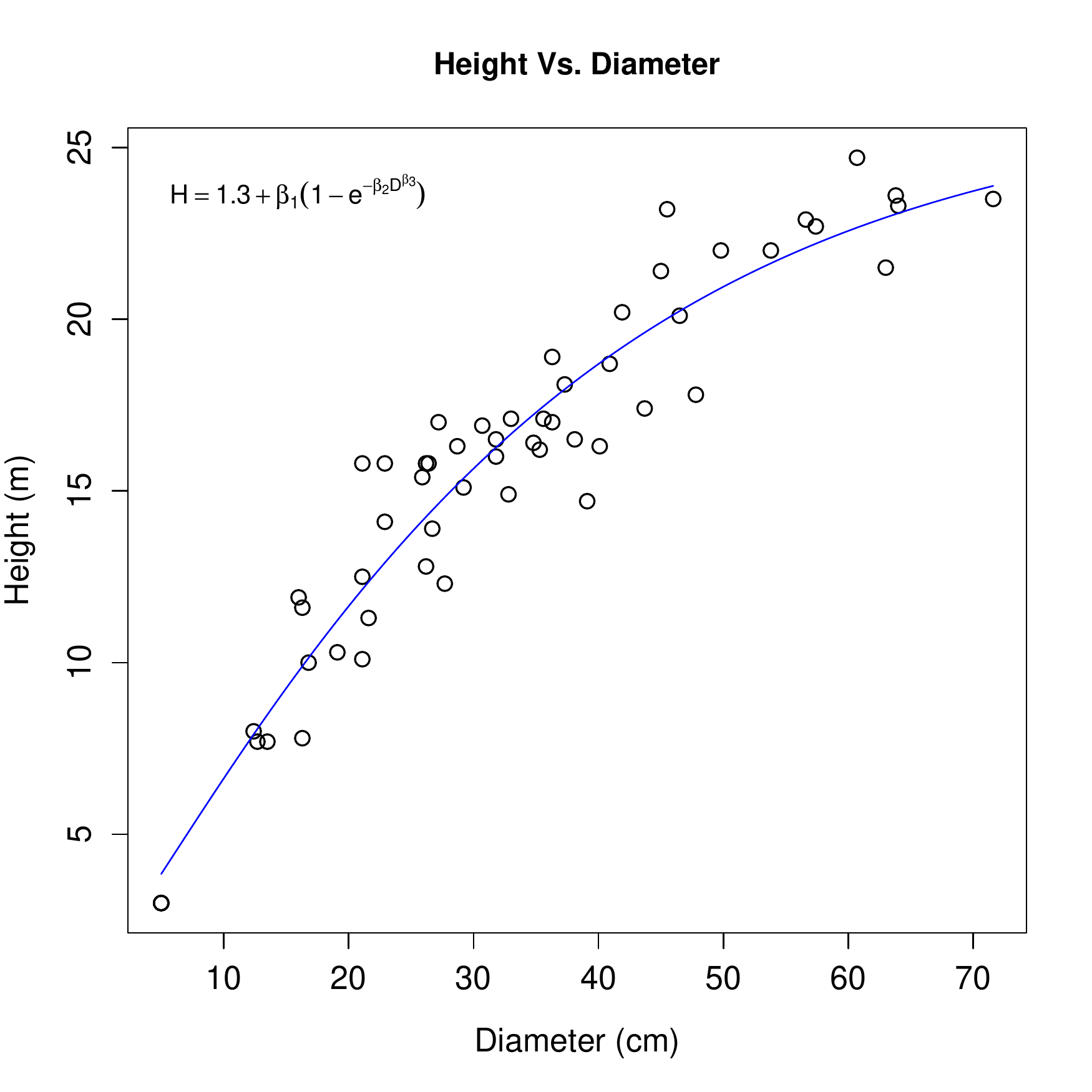} \caption[Weibull growth curve for the height-diameter relationship in plot 55]{Weibull growth curve for the height-diameter relationship in plot 55}\label{fig:growthCurve}
\end{figure}

\begin{kframe}\begin{verbatim}
$summary
         Estimate Std. Error   t value
beta1 25.01192961 2.40070993 10.418555
beta2  0.01670455 0.00492401  3.392468
beta3  1.15619360 0.12410828  9.316007
          Pr(>|t|)
beta1 1.258550e-14
beta2 1.289825e-03
beta3 6.630694e-13

$residuals
 [1]  0.82316850 -0.82069839  3.18177927
 [4] -1.46099717 -0.84691805  2.15198318
 [7] -1.68201648 -0.84691805 -0.36823781
[10] -0.84691805  2.03483956  0.54145978
[13] -0.66930334 -0.13543015 -2.70765800
[16]  0.21749164  1.49299027 -1.05073485
[19]  1.23069666  0.37410634 -1.17243076
[22] -2.09555893 -3.75514371  2.89163490
[25]  1.09509942 -2.42013374  2.37178721
[28] -0.22638929 -0.37990720 -0.84691805
[31] -0.84691805 -1.69128812 -0.26279787
[34]  1.02280885 -0.26430133  1.47743361
[37]  1.55485800 -2.51596865  0.14246603
[40] -0.13947040 -0.53794201 -2.20953508
[43]  1.11600327  0.23720213 -1.44514200
[46]  0.36843804  0.44412866  1.19163490
[49]  1.27175765 -0.92878353 -0.78468991
[52] -0.36241051  1.00941474  0.50189082
[55] -2.03156196  3.66843804  1.70444107
[58]  0.08936654

$`var-cov`
             beta1         beta2
beta1  2.374618630  3.255174e-03
beta2  0.003255174  9.989699e-06
beta3 -0.105817382 -2.393425e-04
              beta3
beta1 -0.1058173816
beta2 -0.0002393425
beta3  0.0063462334

$`residual Std. Error`
[1] 1.557911
\end{verbatim}
\end{kframe}
\end{knitrout}

The output list consists of five parts:

\begin{enumerate}
\item \code{summary}: estimated parameters, standard error of the estimators, $t$-value and corresponding $p$-value.
\item \code{residuals}: residuals of the fitted growth curve.
\item \code{`var-cov`}: variance-covariance matrix of the model coefficient estimators.
\item \code{`residual Std. Error`}: residual standard error.
\item A scatterplot of the height-diameter relationship with the fitted growth curve (Figure~\ref{fig:growthCurve})
\end{enumerate}

\section{Future development}

We are currently working on software functionality that includes: 1) a broader set of goodness-of-fit measures for the Bayesian models; 2) a function that automates ``optimal'' model/method selection that searches among all methods in Table~\ref{tabMethods} with selection based on a set of user identified goodness-of-fit measures; 3) expanded set of mixture models, e.g., \cite{venturini2008}; 4) addition of random effects that accommodate grouped distribution and height-diameter data, e.g., stands, species, functional types, etc.

\section{Conclusion}

A common task in forestry is modeling the diameter distribution of trees in a given forest stand. In this work, we have developed and introduced an \proglang{R} package called \pkg{ForestFit} that provides numerous models and estimation techniques for modeling diameter distributions from both individual tree data and grouped data, as well as providing additional functionality for simulating finite mixture distributions and fitting growth curves to height-diameter data. While this package was developed in the context of forestry, the models we fit and simulate have numerous applications throughout other ecological and environmental fields, such as botany, fisheries, and hydrology. For example, the JSB distribution is used in hydrology to model rain drop size distribution \cite{cugerone2015johnson} and finite mixture distributions are broadly applied in environmental and ecological statistics (for numerous examples, see \cite{hooten}), suggesting \pkg{ForestFit} can have wide use across ecological and environmental fields. \pkg{ForestFit} is freely available on CRAN at \url{https://cran.r-project.org/web/packages/ForestFit/index.html}.


\appendix

\section{Probability density functions used in \texttt{ForestFit}}\label{apa}
\begin{itemize}
\item BS ${f(x,\theta)=\frac{\sqrt{\frac{x-\mu}{\beta}}+\sqrt{\frac{\beta}{x-\mu}}}{2\alpha (x-\mu)}\phi \biggl( \frac{\sqrt{\frac{x-\mu}{\beta}}-\sqrt{\frac{\beta}{x-\mu}}}{\alpha}\biggr),}$

\item Burr XII ${f(x,\theta)=\alpha \beta x^{\alpha-1} \bigl(1+x^{\alpha}\bigr)^{-\beta-1},}$

\item Chen ${f(x,\theta)=\alpha \beta x^{\alpha} \exp\biggl\{x^\alpha-\beta \exp\bigl\{x^\alpha\bigr\}+\beta\biggr\},}$

\item Fisher ${f(x,\theta)=\frac{\Gamma\bigl(\frac{\alpha+\beta}{2}\bigl)}{\Gamma\bigl(\frac{\alpha}{2}\bigl) \Gamma\bigl(\frac{\beta}{2}\bigl)}\Bigl( \frac{\alpha}{\beta}\Bigl)^{\frac{\alpha}{2}} x^{\frac{\alpha}{2}-1}\Bigl[1+\frac{\alpha}{\beta}x\Bigr]^{-\frac{\alpha+\beta}{2}},}$

\item Fr\'{e}chet ${f(x,\theta)=\frac{\alpha}{ \beta} \Bigl( \frac {x}{\beta}\Bigr) ^{-\alpha-1}\exp\Bigl\{ -\Bigl( \frac {x}{\beta}\Bigr)^{-\alpha} \Bigr\},}$

\item gamma ${f(x,\theta)=\frac{ x^{\alpha-1} }{\beta^\alpha \Gamma(\alpha)}\exp\Bigl\{-\frac {x}{\beta}\Bigr\},}$

\item GE ${f(x,\theta)=\beta\exp\bigl\{-\beta (x-\mu)\bigr\} \Bigl[1-\exp\bigl\{-\beta (x-\mu)\bigr\}\Bigr]^{\alpha-1},}$

\item Gompertz ${f(x,\theta)=\beta\exp\bigl\{\alpha x\bigr\} \exp\Bigl\{\frac{\beta \exp\{\alpha x\}-1}{\alpha} \Bigr\},}$

\item Johnson's SB $f(x,\theta) = \frac {\delta \lambda\exp\biggl\{-\frac{1}{2}\Big[\gamma+\delta\log \Bigl(\frac{x-\xi}{\lambda+\xi-x}\Bigr) \Big]^2\biggr\}}{\sqrt{2\pi}(x-\xi)(\lambda+\xi-x)},$

\item log-logistic ${f(x,\theta)=\frac{ \alpha}{ \beta^{\alpha}} x^{\alpha-1} \Bigl[ \Bigl( \frac {x}{\beta}\Bigr)^\alpha +1\Bigr]^{-2},}$

\item log-normal ${f(x,\theta)=\frac{\exp\Bigl\{ -\frac {1}{2}\Bigl( \frac {\log x-\alpha}{\beta}\Bigr) ^2\Bigr\}}{\sqrt{2\pi}
 \beta x },}$

\item Lomax ${f(x,\theta)=\frac{\alpha \beta}{(1+\alpha x)^{\beta+1}},}$

\item skew-normal ${f(x,\theta)=2\phi\Bigl(\frac{x-\alpha}{\beta}\Bigr)\Phi\Bigl(\lambda\frac{x-\alpha}{\beta}\Bigr),}$

\item Weibull ${f(x,\theta)=\frac {\alpha}{\beta} \Bigl( \frac {x-\mu}{\beta} \Bigr)^{\alpha - 1}\exp\Bigl\{ -\Bigl( \frac {x-\mu}{\beta}\Bigr)^\alpha \Bigr\},}$
\end{itemize}
where $\theta$ is the family parameter vector; $\phi(\cdot)$ and $\Phi(\cdot)$ denote the density function and distribution function of the standard normal distribution, respectively. Noticably, the three-parameter BS, gamma, GE, and Weibull distributions simplify to the two-parameter BS, gamma, GE, and Weibull distributions, respectively, when $\mu = 0$.

\printcredits
\bibliographystyle{cas-model2-names.bst}
\bibliography{ref}

\end{document}